# Kelly's Criterion in Portfolio Optimization: A Decoupled Problem

Zachariah Peterson[1,*]

[1]Adams State University
School of Business
208 Edgemont Blvd.
Alamosa, CO 81101
petersonz1@grizzlies.adams.edu

* Correspondence: petersonz1@grizzlies.adams.edu

**Abstract:** Kelly's Criterion is well known among gamblers and investors as a method for maximizing the returns one would expect to observe over long periods of betting or investing. This paper will show how Kelly's Criterion can be incorporated into standard portfolio optimization models that include a risk function. The model developed here combines the risk and return functions into a single objective function using a risk parameter. This model is then solved for a portfolio of 10 stocks from a major stock exchange using a differential evolution algorithm. Monte Carlo calculations are used to directly simulate and compare the average returns from the Mean Variance and Kelly portfolios. The results show that Kelly's Criterion can be used to calculate optimal returns and can generate portfolios that are similar to results from the Mean Variance model. The results also show that evolutionary algorithms can be successfully applied to solve this unique portfolio optimization problem.

**Keywords:** portfolio optimization; Kelly criterion; differential evolution; mean variance; logarithmic utility

**Key Messages:**

- A decoupled form of logarithmic utility can be used to optimize portfolios

- The decoupled Kelly model generates portfolios with similar returns as the MV model

- The results were reached using test data with an intuitive solution

## 1. Introduction

In general, portfolio optimization problems aim to determine an optimal allocation of wealth among a pool of candidate securities. Portfolio optimization was first discussed in 1952 by Harry Markowitz in his work on modern portfolio theory (MPT) (Markowitz 1952). According to MPT, an optimum portfolio can be arranged such that return is maximized for a specified level of risk, or vice-versa, where risk is minimized for a specified level of return. Many formulations of portfolio optimization problems are linear or quadratic, depending on the definition of portfolio risk that is used in the particular problem. The original formulation of Markowitz is known as the Mean Variance (MV) model and treats return on a portfolio of investments using historical averages of changes in market prices for each asset. The total portfolio return was defined to be a weighted sum of returns from individual investments. Risk was defined as variance of returns and is found by taking the inner product of the covariance matrix for the assets in the portfolio. Later models for asset pricing, such as the Capital Asset Pricing Model (CAPM) (Fama and French 2004), would continue to use the covariance among changes in asset values to quantify risk. Other models, for example in (Bichpuriya and Soman 2016; El Ghaoui, et. al. 2003) use value-at-risk (VaR) or conditional value-at-risk (CVaR) to model the variation in portfolio returns.

The MV approach to portfolio optimization, where returns are defined using average changes in market prices of assets over time, over-simplifies the problem. A better reflection of reality is to determine the probability distribution of price changes for each of the assets in the portfolio and reformulate the return function in terms of these probabilities. This is done in (Yang and Liu 2016), where returns are treated as fuzzy numbers. Portfolio optimization based on MPT has also been used in electricity generation and distribution, where electricity demand is treated as a random variable (Bichpuriya and Soman 2016). Both of these examples formulate the portfolio return function in terms of expectation values and covariant risk, generating a linear (Bichpuriya and Soman 2016; El Ghaoui, et. al. 2003) or quadratic (Yang and Liu 2016) objective function problem that has the same form as the MV formulation. The linear return function in the MV model has been used by many authors in the automation literature (Bichpuriya and Soman 2016; El Ghaoui, et. al. 2003; Yang and Liu 2016; Kamili and Riffi 2016; Chen, et. al. 2012; Zaheer and Pant 2016; Korczak and Roger 2000; Ma, et. al. 2012; Chang, et. al. 2009) and is treated as something of a standard model for portfolio optimization. However, one can show from probability theory that the optimum return on an investment (or portfolio of investments) is not a linear function of the fraction of wealth placed in each investment.

Kelly's Criterion is well known among gamblers as a betting strategy (Rotando and Thorpe 1992; Browne and Whitt 1996; Thorpe 1997). Kelly's result is, in its simplest sense, a solution to an optimization problem which maximizes logarithmic utility and was originally applied to a technical problem in information theory (Kelly 1956; Kim 2008). The Kelly Criterion has been discussed in contexts outside of gambling, for example, in engineering economics (Kim 2008). These ideas were later embraced by gamblers and were used to maximize the winnings one would expect to see from a large number of bets with well-defined probabilities. A gambler that bets a critical fraction of their wealth per bet can expect to see an average rate of return per bet that maximizes their total return over time (Rotando and Thorpe 1992; Browne and Whitt 1996; Thorpe 1997; Nekrasov 2014; Kim 2008; Vince 2011). In its simplest form, where the outcome of each bet was considered binary with well-defined odds and probabilities, and successive bets are mutually independent, Kelly's result is a simple formula for the critical fraction of wealth that will maximize a gambler's average return over a large number of bets (Rotando and Thorpe 1992; Browne and Whitt 1996; Thorpe 1997; Vince 2011; Nekrasov 2014). More complicated bets do not have such simple formulae. However, one can show in general that Kelly's return function is concave (Rotando and Thorpe 1992; Vince 2011) and a maximum solution exists. Although this was originally applied to a set of discrete i.i.d outcomes using the Central Limit Theorem, it has also been applied to continuously distributed random variables via analogous limit theorems (Rotando and Thorpe 1992; Browne and Whitt 1996; Nekrasov 2014).

In the applied mathematics literature, Kelly's Criterion has been applied to investing in a single asset for discretely (Browne and Whitt 1996; Thorpe 1997) and continuously distributed (Merton 1971; Rotando and Thorpe 1992; Browne and Whitt 1996; Thorpe 1997; Pestein and Sudderth 1985) changes in asset values. Kelly's Criterion has also been applied to a single asset using a Bayesian formulation of continuously distributed outcomes (Browne and Whitt 1996) in discrete and continuous time. In all cases, one finds the rate of return depends nonlinearly on the fraction of one's wealth that is placed on successive bets. All of these previous applications address a portfolio containing a single risky asset. One case (Browne and Whitt 1996; Merton 1971) addressed a portfolio of one risky asset and one riskless bond. The Kelly Criterion has been applied to a portfolio by considering a considering a continuous probability distribution for changes in the S&P 500 index (Rotando and Thorpe 1992), however it was not addressed how one should distribute capital among the stocks that comprise the index. Progress was made in this area more recently (Nekrasov 2014); the Kelly Criterion has been applied to a portfolio of 7 risky stocks, with the goal of optimizing return from the entire portfolio. However, one has yet to see the Kelly Criterion extended to a portfolio optimization problem that considers simultaneous minimization of portfolio risk.

Solving this type of optimization problem requires an efficient algorithm for nonlinear objective functions in any dimension. One class of algorithms that is applicable to this type of problem are meta-heuristics. A subclass of these methods comprises evolutionary algorithms, which are intended to mimic the behavior of natural systems and are based on stochastic search methods. Various evolutionary algorithms have been applied to optimization problems in a wide range of fields including circuit design, mechanical engineering (Storn and Price 1997), aerodynamics (Rogalsky, et. al. 1999), and medical imaging (Kamili and Riffi 2016; Qin, et. al. 2009). Several meta-heuristics (swarm optimization (Kamili and Riffi 2016; Chen, et. al. 2012), differential evolution (Zaheer and Pant 2016; Korczak and Roger 2000; Ma, et. al. 2012), and genetic algorithms (El Ghaoui; et. al. 2003; Chang, et. al. 2009)) have been used to solve portfolio optimization problems as well.

The intention of this paper is to examine a general case where changes in values of multiple assets are correlated and continuously distributed, and where one can place fractions of capital in a large number of investments. It will be shown here that one can use Kelly's Criterion to define a multidimensional nonlinear function for the rate of return from a portfolio of these investments. The return function in portfolio optimization will be reformulated using Kelly's Criterion in discrete time and solved using differential evolution. The theory will demonstrate the existence of two types of return functions, which are termed the decoupled and the coupled return functions. By following Kelly's process, the decoupled return function for a portfolio of $N$ investments can be shown to be a nonlinear function of the fraction of total wealth that is placed in each investment. The result transforms the risk function from a second degree to a fourth degree polynomial objective function in $N$ dimensions. The results will show that the use of the decoupled return function results in comparable or better portfolios than the MV return function. The results also show that differential evolution is a viable algorithm for solving portfolio optimization models based on Kelly's Criterion.

The second section will derive the nonlinear return function for a portfolio of assets with specified distributions of changes in asset values. This is then used to calculate the average return and the variance of returns. The third section discusses the incorporation of these results into some well-known models for portfolio optimization. A cardinality-constrained portfolio optimization problem that combines objective functions with a risk parameter will be presented. The fourth section will present and discuss numerical solutions obtained using differential evolution. The results obtained using the decoupled return function will be compared to the results obtained using the risk and return functions in the MV model under the same set of constraints. The fifth section will present and discuss conclusions, as well as offer a guide to further research in this area.

**2. The nonlinear return function**

To understand the necessity of reformulating the return function in portfolio optimization model based on mean-variance models, the original linear return function of Markowitz will first be examined. The linear return function defines the return from a portfolio of $N$ investments with initial value of $W_0$ as a sum of returns from each of the investments $r_i$, with each investment return weighted by the fraction of the portfolio that is invested in each asset $f_i$. Returns on investments are random variables and the value of a portfolio of $N$ investments after $n$ accrual periods is an $n$-fold product of accruals, where each accrual multiplies the value of the $i$th investment by a factor $1 + X_{i,j}$. $X_{i,j}$ is a random variable representing the change in value of the $i$th from period $j-1$ to $j$. Here $X_{i,j}$ could be any type of investment whose value can be tracked over time. If, for example, $X_{i,j}$ represents a stock, then $X_{i,j} = \left(\frac{S_{i,j+1}}{S_{i,j}}\right) - 1$, where $S_{i,j}$ is the value of the $i$th stock at period $j$. The value of the portfolio after the $n$th period is taken to be

$$W_n = \sum_{i=1}^{N} F_i W_0 \left(1 + X_{i,j}\right)^n. \tag{1}$$

In the MV model, the average return rate after a single period is given by the sum of expected values of the fractional changes in asset values as

$$R_{avg} = \sum_{i=1}^{N} F_i E[X_i]. \tag{2}$$

Markowitz's MV model, as well as others working in this area (Bichpuriya and Soman 2016; El Ghaoui, et. al. 2003; Yang and Liu 2016; Kamili and Riffi 2016; Chen, et. al. 2012; Zaheer and Pant 2016; Korczak and Roger 2000; Ma, et. al. 2012; Chang, et. al. 2009), have formulated each of the $E[X_{i,j}]$ values as an average over a large number of changes in asset values over discrete time periods. These changes can occur daily, weekly, etc. In some cases, particularly in (Bichpuriya and Soman 2016; Yang and Liu 2016) this is done by first determining a probability density function for changes in asset values between successive points in time and then calculating the value of $E[r_i]$ directly from this distribution. Here, no assumption has been made on the type of investments in the portfolio; the only assumption is that each of the distribution functions for asset changes are known a priori or can be determined from market data.

Kelly's formulation can now be used to define the portfolio's return function in terms of the expected value of an exponential rate constant. Here we can derive the form of this expectation value directly from a formulation of the return function for the portfolio. This can be done by chopping the initial investment capital $W_0$ into $N$ fractions of size $f_i W_0$ and applying Kelly's Criterion to each portion of capital.

Here we will employ the "fixed-fractional" strategy (Rotando and Thorpe 1992; Browne and Whitt 1996), where the fraction of wealth invested is kept fixed over the entire investment horizon. When compared to other investing strategies, this strategy of investing a fixed proportion of wealth in a risky asset such that return rate of the portfolio is maximized will also minimize the time it takes for the portfolio to reach a target value. This has been proven in both discrete and continuous time (Rotando and Thorpe 1992; Pestein and Sudderth 1985). As the values of assets in the portfolio change over time, capital may need to be moved between assets such that the fractions of capital placed in each asset are restored to their initial values.

Let the total value of the portfolio at period $j$ be $W_j$, and let the portfolio value due to returns from the $i^{th}$ asset at period $j$ be $w_{i,j}$. Initially, we consider the case of a single asset, and then expand the return function to multiple assets. Applying Kelly's Criterion under the fixed-fractional strategy requires we invest some fraction $f_i$ of this amount into a risky security at $j = 0$ such that we optimize the return rate. In terms of the fractional changes in asset values $X_{i,j}$, then the value of the portfolio invested in the $i^{th}$ security over a single period is

$$w_{i,j+1} = W_j - f_i W_j + f_i W_j (1 + X_{i,j}). \tag{3}$$

At period $j$, an amount of capital equal to $w_{i,j} = f_i W_j$ is invested and the investment appreciates in value to $f_i w_{i,j}(1 + X_{i,j})$ from period $j$ to $j + 1$, leaving $W_j - f_i W_j$ to invest in other assets. For the first accrual, (3) reduces to

$$w_{i,1} = W_0 - f_i W_0 + f_i W_0 (1 + X_{i,0}) = W_0 (1 + f_i X_{i,0}). \tag{4}$$

Using (3) and (4), the value of the portfolio after $n$ periods can be found by induction using permutations on index $j$ in (4), followed by a summation over index $i$. Iterating index $j$ from $0$ to $n$ and using the Binomial Theorem, we have the following equation for the value of the portfolio due to returns from the $i^{th}$ investment after $n$ periods:

$$w_{i,n} = W_0 \prod_{j=1}^{n}(1 + f_i X_{i,j}). \tag{5}$$

Under the fixed-fractional investing strategy, the fraction of the portfolio invested in asset $i$ is $f_i w_{i,j} \, \forall \, j$. Therefore the total portfolio value at period $j$ is $W_j = \sum_{i=1}^{N} f_i w_{i,j}$. The total portfolio value after $n$ periods is

$$W_n = \sum_{i=1}^{N} f_i W_0 \prod_{j=1}^{n}(1 + f_i X_{i,j}). \tag{6}$$

From (6) we see that the portfolio value after $n$ periods depends nonlinearly on the fraction of total wealth that is placed in each investment. The rate of return is

$$R_{avg} = \sum_{i=1}^{N} f_i \prod_{j=1}^{n}(1 + f_i X_{i,j}) - 1. \tag{7}$$

Equation (7) is the return function for the portfolio after $n$ accrual periods. We now have a function that treats returns as draws from a set of correlated random variables, and this function can be used to calculate the expected return and the variance in returns (i.e. portfolio risk). In (7) the actual fraction of wealth that is invested in the $i^{\text{th}}$ asset is equal to $f_i^2$.

Now that returns are defined in terms of the outcomes from a set of random variables, it is of interest to know the expected value of the return function (7). This defines the expected return from the number of accruals $n$ becomes large, i.e. as the process becomes continuous. Applying a logarithmic identity to the return function allows the products in (7) to be converted to a sum. This converts each of the $(1 + f_i X_{i,j})$ product terms in the series to an exponential function of a summation:

$$\prod_{j=1}^{n}(1 + f_i X_{i,j}) = \exp\left(\sum_{j=1}^{n} \ln(1 + f_i X_{i,j})\right). \tag{8}$$

Here a natural logarithm has been used, but one can use any base for the logarithm; Kelly's original paper used a base of 2 (Kelly 1956). In general, (8) represents outcomes over multiple time steps where returns over successive time steps may be correlated. In proceeding further, the multi-step problem in (8) will be solved by considering returns that accrue over a single time step.

To proceed further requires the Central Limit Theorem, which states that the average of a large number of outcomes from a family of i.i.d. random variables converges to the expected value of the random variable. If we assume the parameters (drift and volatility) defining the distribution of $X_{i,j}$ are constant over the entire investment horizon, then each of the $X_{i,j}$ terms in (8) are identically distributed outcomes over the entire investment horizon, i.e. $\forall \, j$. The assumption that successive returns for a single asset over multiple time steps are independent is common in models for asset evolution (Joshi 2008), and there is evidence to support the idea that successive changes in individual stock values behave like i.i.d. random variables (Moore 1962; Rotando and Thorpe 1992). Thus it is reasonable to assume that the $X_{i,j}$ terms are also independent $\forall \, j$, and the Central Limit Theorem can be applied as a first approximation. In general, return rates may be correlated over time, i.e. returns may be path-dependent. It is important to note that if successive returns on a single asset are strongly correlated, a different limit theorem would need to be applied. Assuming successive returns for an individual asset are uncorrelated, the Central Limit Theorem is used here to convert the summation in the argument of the exponential function in (8) to an expected value multiplied by the total number of accruals $n$:

$$\prod_{j=1}^{n}(1 + f_i X_{i,j}) \sim \exp(nE[\ln(1 + f_i X_i)]) \tag{9}$$

where $X_i$ is an accrual over a single period. Equation (9) expresses the limit as the number of periods becomes large and is an approximation; this has been discussed previously (Samuelson 1971). This is accurate to within a certain confidence interval that depends on the square root of the number of accruals. As the number of periods in the investment horizon increases the expected value grows faster than the size of the confidence interval, and the actual return approaches the expected return in (9) with probability 1. In other words, the predictions under Kelly's Criterion are asymptotic (Samuelson 1971; Rotando and Thorpe 1992; Thorpe 1997).

The average return for a single accrual can be found by approximating the summation in (8) by the mean. Placing the approximation in (9) back into (7) and gives the average return for the $i^{th}$ asset over a single time step:

$$R_{avg} = \sum_{i=1}^{N} f_i \exp(E[\ln(1 + f_i X_i)]) - 1. \tag{10}$$

Equation (10) will henceforth be referred to as the *decoupled return function* for reasons that will soon be apparent.

It is important to note that the same process used to derive the decoupled return function above could have been applied to (1) immediately without splitting the portfolio into $N$ fractions. If we apply this process to (1), we arrive at a return function that couples the random variables and the wealth fractions $F_i$ into a single expectation value in $N$ dimensions. This result for the average return over a single period can be called the *coupled return function* and is given by

$$R_c = \exp(E[\ln(1 + \sum_{i=1}^{N} F_i X_i)]) - 1. \tag{11}$$

This equation uses the same assumptions used to derive the decoupled return function, i.e. that successive returns for each asset over each accrual period are i.i.d. random variables (path-independent). In the general case, this expectation value is an $N$-dimensional integral with each of the $F_i$ values appearing as parameters. In both return functions, the global optimum for return on investment may occur when $\sum_{i=1}^{N} F_i \neq 1$. A version of this equation was previously derived (Browne and Whitt 1996) to optimize a portfolio of one risky and one riskless asset. A portfolio of 7 risky assets has also been treated in the literature (Nekrasov 2014). In both cases, simultaneous minimization of the covariant risk function was not incorporated in the optimization process.

Both return functions require a priori knowledge of the joint distribution function $p(\mathbf{X})$ for the assets in the portfolio. As equation (11) involves an $N$-dimensional integral that is not seperable, it is computationally more complex and may not be analytically solvable (depending on the form of $p(\mathbf{X})$), even in the case where the $X_i$ random variables are independent. In contrast, the decoupled return function (10) will reduce to a marginal distribution for a single asset $p(X_i)$ as the dependence of $N-1$ of the variables is eliminated via integration over the space of outcomes in $N-1$ dimensions. Thus the decoupled problem may be preferable both analytically and numerically. If the time required to compute a single expectation value in the decoupled problem is $T$, then the time required to compute all $N$ expectation values is $NT$. The time required to compute the expectation value in the coupled problem is $T^N$, and the decoupled problem is clearly computationally more efficient. The coupled return function in (11) has been optimized using a so-called "Monte Carlo grope algorithm" that is very similar to differential evolution (Nekrasov 2014). In this case the evaluation time is $Nn_s T_s$, where $n_s$ and $T_s$ are the number of samples and time per sample respectively. The remainder of this paper will focus on solving the portfolio optimization problem using the decoupled return function.

Equation (7) is a linear combination of random variables and the variance of this function defines the portfolio risk (Markowitz 1952). The variance of a linear combination of random variables can be written as an inner product of the covariance matrix for these random variables (Fisher 1990). Specifically, let $Z$ be a linear combination of random variables with covariance matrix $[M]$, and let $[a]$ be the column vector defining the coefficients. The variance of $Z$ is given by the following

$$Var[Z] = [a]'[M][a]. \qquad (12)$$

In (12), $[a]'$ is the transpose of $[a]$. Using (12) we can calculate the variance of a single return (i.e. when $n = 1$ in (7)). The random variables in the return function are $1 + f_i X_i$, and the coefficients are $a_i = f_i$. The entries in the covariance matrix are

$$M_{ij} = Cov[X_i, X_j]. \qquad (13)$$

The coefficient vector in (7) is $[a] = [f^2]$. From the inner product of the covariance matrix defined by (13), the risk function for the portfolio is

$$Var[R] = \sum_{i=1}^{N}(f_i^4 M_{ii} + 2\sum_{j>i}^{N} f_i^2 f_j^2 M_{ij}). \qquad (14)$$

## 3. The decoupled return function in portfolio optimization

Equations (10) and (14) form a dual objective optimization problem. However, the two results can be combined into a single objective function using a risk parameter. In this model, the risk parameter $P$ measures an investor's risk indifference (large $P$) or risk aversion (small $P$) and allows a portfolio to be tailored to an individual investor's risk preferences. In the cardinality-constrained efficient frontier (CCEF) model (Kamili and Riffi 2016), the risk parameter takes values in the interval $[0, 1]$; these are the bounds that will be used in the model that will be presented below. It should be noted that these are not the only bounds that have been used in models based on risk parameters. The CVaR model for portfolio optimization in (Bichpuriya and Soman 2016), for example, used bounds of $[0, 3]$ for the risk parameter in their single objective problem. To form the objective function for portfolio optimization, the return and risk equations from the MV model are first combined into a single objective problem, then a second model will be presented that combines (10) and (14) into a single objective problem.

The imposition of cardinality-constraints in portfolio optimization limits the wealth fractions invested in different assets within some specified range. Formally, this type of constraint can be written using a pair of vectors $[K_{min}]$ and $[K_{max}]$. If we form the wealth fractions into a vector $[f]$, the boundary condition in the decoupled Kelly problem is

$$[K_{min}] \leq [f] \leq [K_{max}]. \qquad (15)$$

The sum of components of the $[K_{max}]$ vector must be less than the total number of assets $N$, which is equivalent to limiting $K_{max,i} < 1$ for all assets.

The equations in the MV model for risk and return form the following cardinality-constrained portfolio optimization model:

$$\max P \sum_{i=1}^{N} F_i E[r_i] - (1-P) \sum_{i=1}^{N}\left(F_i^2 M_{ii} + 2\sum_{j>i}^{N} F_i F_j M_{ij}\right) \qquad (16)$$

$$\text{subject to} \sum_{i=1}^{N} F_i = 1$$

$$K_{min,i} \leq F_i \leq K_{max,i} \; \forall \; i.$$

Here, the $[M]$ matrix appears in (13), $P$ is the risk parameter bounded in the interval $[0, 1]$. The $E[X_i]$ terms are calculated directly from market data. This is a similar formulation to the model from Kamili and Riffi (2016) and uses the same risk and return terms.

The above model is now reformulated using the decoupled return and risk functions given by (10) and (14) respectively. The portfolio optimization problem incorporating Kelly's Criterion is given by the following:

$$\max P \sum_{i=1}^{N} f_i(\exp(E[\ln(1 + f_i X_i)]) - 1) - (1 - P) \sum_{i=1}^{N} \left( f_i^4 M_{ii} + 2 \sum_{j>i}^{N} f_i^2 f_j^2 M_{ij} \right)$$

$$\text{subject to} \sum_{i=1}^{N} f_i^2 = 1 \tag{17}$$

$$K_{min,i} \leq f_i^2 \leq K_{max,i} \; \forall \; i.$$

This model will henceforth be called the *decoupled Kelly model*. Here, the optimization variables are the set of $\{f_i\}$, and the actual wealth fraction invested in each asset is the set of $\{f_i^2\}$.

The individual terms in (10) are known to be concave functions (Rotando and Thorpe 1992; Vince 2011), and therefore the linear combination of these terms with positive coefficients is also a concave function. Thus a solution exists that will maximize (10). Each term in (14) is an even degree polynomial function, and asset pairs with positive correlation ($M_{ij} > 0$) are convex. The linear combination of these convex terms is also convex and a solution exists that will minimize equation (14). These two functions are combined in equation (17) via linear combination, where the $-(1 - P)$ coefficient changes the convex terms in the risk function to concave terms. Although negatively correlated asset pairs ($M_{ij} < 0$) will then be transformed to convex functions when multiplied by the $-(1 - P)$ coefficient, and these terms will increase the value of the objective function. One can conclude that a solution must exist that will maximize the portfolio optimization problem in equation (17).

It has been noted in the literature that wealth fractions generated from Kelly's Criterion are only true wealth fractions under certain conditions (Vince 2011). This requires that the probability of total loss of investment in an asset be nonzero over the entire course of the investment horizon. If this condition is not met, the Kelly Criterion may return optimum $f_i$ values that are greater than 1 (Rotando and Thorpe 1997; Kim 2008; Vince 2011). A specific example was shown by Rotando and Thorpe (1997); they considered a truncated normal distribution for returns on the S&P index, i.e. the probability of total loss was defined to be zero. Their analysis returned the result $F = 1.69$. A wealth fraction value under the Kelly models can be interpreted as a "leveraging factor" (Vince 2011) rather than as a fraction of investment capital.

For the decoupled Kelly model, one must first determine the drift and volatility parameters using the solution to the stochastic differential equation for geometric Brownian motion for each of the assets in the portfolio. The stochastic differential equation for the $i^{\text{th}}$ stock is defined as an Ito process given by

$$dS_t = S_t(\mu dt + \sigma dW). \tag{18}$$

The solution to this stochastic differential equation with constant coefficients is the well-known log-normal distribution for changes in stock values over a single time period (Joshi 2008). The $X_i$ random variable as a function of time in (17) is defined as

$$X_i(t + \Delta t) = \frac{S_i(t+\Delta t)}{S_i(t)} - 1 = e^{\left(\mu_i - \frac{1}{2}\sigma_i^2\right)\Delta t + \sigma_i\sqrt{\Delta t}y} - 1, \tag{19}$$

where $y$ is a standard normally-distributed random variable that defines the magnitude of an up or down movement of the stock value over the period $\Delta t$, and $\Delta t = 1$ month.

Equation (19) can now be used to relate the drift and volatility to the average value and sample variance of monthly returns from the market data (see table 1 in (Zaheer and Pant 2016)). Taking the expected value and variance of (19) yields two formulas for the drift and volatility of the $i^{th}$ stock over a single month. Let $Avg[R]$ and $Var[R]$ be the sample mean and sample variance calculated from return data respectively. Taking the expected value and variance of (19) gives the following equations for the drift and volatility of the $i^{th}$ stock over a single month:

$$\mu_i = \ln(1 + Avg[R])$$
$$\sigma_i = (\ln(Var[R]e^{-2\mu_i} + 1))^{\frac{1}{2}} \tag{20}$$

The values for drift, volatility, and the sample covariance matrix are summarized in tables 1 and 2 below. Table 1 also shows the values of the first-to-second moment ratios for each stock calculated directly from equation (19).

Table 1. Parameters for the portfolio optimization models in (16) and (17).

|  | $X_1$ | $X_2$ | $X_3$ | $X_4$ | $X_5$ | $X_6$ | $X_7$ | $X_8$ | $X_9$ | $X_{10}$ |
|---|---|---|---|---|---|---|---|---|---|---|
| $Avg[R]$ | 0.1750 | 0.0995 | 0.3398 | 0.2366 | 0.1149 | 0.2799 | 0.2158 | 0.2593 | 0.2686 | 0.4405 |
| $\mu_i$ | 0.1613 | 0.0949 | 0.2925 | 0.2123 | 0.1087 | 0.2468 | 0.1954 | 0.2305 | 0.2379 | 0.3650 |
| $\sigma_i$ | 0.3516 | 0.3277 | 0.3804 | 0.2660 | 0.2205 | 0.4398 | 0.2750 | 0.2167 | 0.3477 | 0.1991 |
| $\frac{E[X]}{E[X^2]}$ | 1.5521 | 1.6268 | 1.5447 | 1.0347 | 0.8814 | 1.9151 | 1.0856 | 0.8129 | 1.4232 | 0.7285 |

Table 2. Sample covariance matrix from the data in (Zaheer and Pant 2016).

|  | $X_1$ | $X_2$ | $X_3$ | $X_4$ | $X_5$ | $X_6$ | $X_7$ | $X_8$ | $X_9$ | $X_{10}$ |
|---|---|---|---|---|---|---|---|---|---|---|
| $X_1$ | 0.1817 | 0.0978 | 0.1403 | 0.0962 | 0.0481 | 0.1745 | 0.0752 | 0.0574 | 0.1326 | 0.0040 |
| $X_2$ | 0.0978 | 0.1370 | 0.1246 | 0.0696 | 0.0288 | 0.1823 | 0.0924 | 0.0438 | 0.1121 | 0.0402 |
| $X_3$ | 0.1403 | 0.1246 | 0.2778 | 0.1352 | 0.0582 | 0.1756 | 0.0914 | 0.0953 | 0.1621 | 0.0703 |
| $X_4$ | 0.0962 | 0.0696 | 0.1352 | 0.1121 | 0.0443 | 0.1440 | 0.0617 | 0.0572 | 0.1198 | 0.0321 |
| $X_5$ | 0.0481 | 0.0288 | 0.0582 | 0.0443 | 0.0619 | 0.0970 | 0.0338 | 0.0537 | 0.0690 | -0.0141 |
| $X_6$ | 0.1745 | 0.1823 | 0.1756 | 0.1440 | 0.0970 | 0.3495 | 0.1543 | 0.1055 | 0.2195 | 0.0291 |
| $X_7$ | 0.0752 | 0.0924 | 0.0914 | 0.0617 | 0.0338 | 0.1543 | 0.1161 | 0.0581 | 0.1160 | 0.0484 |
| $X_8$ | 0.0574 | 0.0438 | 0.0953 | 0.0572 | 0.0537 | 0.1055 | 0.0581 | 0.0763 | 0.0844 | 0.0216 |
| $X_9$ | 0.1326 | 0.1121 | 0.1621 | 0.1198 | 0.0690 | 0.2195 | 0.1160 | 0.0844 | 0.2068 | 0.0466 |
| $X_{10}$ | 0.0040 | 0.0402 | 0.0703 | 0.0321 | -0.0141 | 0.0291 | 0.0484 | 0.0216 | 0.0466 | 0.0839 |

Equations (20) and (21) define a marginal distribution for changes in each of the stock values over a single time period (in this case, over a single month). When taken together with the entries in the covariance matrix, one can define a joint distribution function for changes in the asset values for all 10 stocks in the portfolio. The changes in stock values are jointly normally distributed with the coupling determined by the covariance matrix. However, as was noted previously, the expectation value $E[\ln(1 + f_i X_i)]$ reduces to a one-dimensional integral involving only the marginal distribution. Thus the $i^{\text{th}}$ expectation value in the argument of the exponential function in equations (10) and (17) is given by

$$E[\ln(1 + f_i X_i)] = \frac{1}{\sqrt{2\pi}} \int \ln\left(1 + f_i \left(e^{\mu_i - \frac{1}{2}\sigma_i^2 + \sigma_i y} - 1\right)\right) e^{-\frac{y^2}{2}} dy. \tag{21}$$

There is no analytical solution to this integral, however it can be efficiently evaluated numerically. Here we define $y_j = \frac{\ln\left(\frac{S_{i,j}}{S_{i,0}}\right) - \left(\mu_i - \frac{1}{2}\sigma_i^2\right)}{\sigma_i}$ and use the appropriate differential element for numerical integration.

## 4. Results from differential evolution

The portfolio optimization problems in equations (16) and (17) are solved using differential evolution. This approach is a robust stochastic searching technique that is ideally suited to solving nonlinear problems such as those shown above. The original algorithm for differential evolution is presented in (Storn and Price 1997). The algorithm used here was run on a PC with a 2.4 GHz dual-core processor and 4 GB RAM.

Differential evolution proceeds via three principle steps: mutation, crossover, and greedy selection. During mutation, candidate values for the objective variables in the optimization problem are generated from an initial population. The mutation step generates candidate solutions by combining components from the previously acceptable solution and compares these candidates with the constraints on the variables in the problem. The crossover step randomly incorporates candidates that were generated in the mutation step into a trial solution. The greedy selection step compares the previous acceptable solution with the trial solution. If the trial solution is favorable then it is accepted as the current best solution. Otherwise it is rejected. The algorithm repeats until particular termination criteria have been reached. Here, the stopping criteria are based on the elapsed processing time rather than a fixed number of iterations. The algorithm starts a timer, and once the timer exceeds 30 seconds the algorithm terminates. If the algorithm generates an acceptable new solution during the greedy selection step, the timer resets to zero and the algorithm repeats. The selection process in the mutation step used here was taken from Storn and Price's original C-code in (Storn and Price 1997).

The two critical parameters that control the progress of differential evolution algorithms are the crossover rate $C$ and the scaling factor $F$. The crossover rate is equivalent to the probability that a candidate component generated in the mutation operation crosses over to the trial solution. A value of $C = 0.75$ in all the calculations performed in this study. The scaling factor scales the random combinations generated during the mutation step and is critical to generating new candidate solutions. A number of trial generation strategies can be found in the literature (Storn and Price 1997; Qin, et. al. 2009), all of which use a scaling factor to amplify or diminish mutation. The original formulation of differential evolution restricted $F \in [0, 2]$.

The algorithm used here is a variant on this original formulation (known in the literature as DE/rand/1/bin (Storn and Price 1997)). The generation strategy in DE/rand/1/bin has been shown to be effective in optimizing even degree polynomial and other nonlinear objective functions in many dimensions (Storn and Price 1997; Qin, et. al. 2009). Scaling factors on the order of $F \sim 0.5$ were shown to be effective in solving these problems with nearly 100% reproducibility. Here the value of $F$ is $\frac{1}{2N}\sum_{i=1}^{N}(K_{max,i} - K_{min,i})$. A noise term $N(0, 0.01)$ is also added to the trial solutions generated during the mutation step, where $N(0, 0.01)$ represents a normally-distributed random variable with 0 mean and 0.01 standard deviation. This ensures that there is always some slight perturbation to the solutions that are generated during mutation, particularly in the case where two of the randomly selected assets have very similar values. C-style pseudo-code for this algorithm can be found in the Appendix. Constraint handling for the inequalities in (16) and (17) is accomplished with using the logical scheme set out in Lampinen (2002, equation (9)). The constraint handling has been incorporated into the pseudo-code using ConstraintCheck and ConstraintMod functions. The precision on the equality constraint is set to ±0.01 in the greedy selection step; this allows some flexibility on the selection criteria applied to mutated solutions. Since the total wealth fractions are given by an equality constraint, one of the wealth fractions is selected randomly to be determined by the equality constraint after the first 9 wealth fractions are generated during the mutation step.

Differential evolution algorithms typically start with an initial solution consisting of randomly generated initial values (Qin, et. al. 2009). If one examines the drift and volatility values in table 1, the drift-to-volatility and Sharpe ratios are largest for the 10th stock. One would reasonably expect that the solver will produce solutions that place a large fraction of investment capital in the 10th stock for any value of risk parameter. Previous work in this area has shown that the unconstrained optimal solution for a single asset is $F = \frac{\mu}{\sigma^2}$ in both the discrete (Thorpe 1997; Kim 2008; Vince 2011) and continuous cases (Browne and Whitt 1996). The initial solution is generated such that the solver can reach this intuitive solution and avoid premature convergence. The initial wealth fraction for each stock is set equal to its drift-to-volatility ratio normalized by the sum of drift-to-volatility ratios for all the stocks: $f_{i,(initial)}^2 = \frac{(\mu_i/\sigma_i)}{\sum_{l=1}^{N}(\mu_l/\sigma_l)}$. This type of initialization is useful in preventing premature convergence at a local maximum. The limits from cardinality are set at $K_{min} = 0.05$ and $K_{max} = 0.95$ for all assets.

The portfolio optimization models in equations (16) and (17) were solved for various values of the risk parameter. The solution algorithm was run 20 times in succession for each value of risk parameter so that convergence and sensitivity of the algorithm can be evaluated. The solution generated from a given run is used as the starting point for the following run. Results from the algorithm for each model are shown in the figures below. These figures show how the return and risk values converge over multiple runs for the MV model (figure 1) and the decoupled Kelly model (figure 2). Tables 3 and 4 show the final portfolios generated after the 20th run for each value of risk parameter.

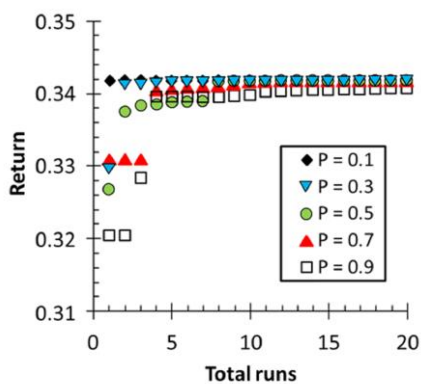

(a)

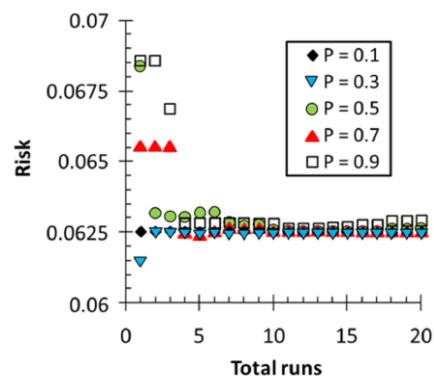

(b)

**Figure 1.** (a) Return and (b) risk values for each run of the solution algorithm for the MV model at various values of risk parameter $P$.

Table 3. Final portfolios generated for the MV model.

|  | $F_1$ | $F_2$ | $F_3$ | $F_4$ | $F_5$ | $F_6$ | $F_7$ | $F_8$ | $F_9$ | $F_{10}$ |
|---|---|---|---|---|---|---|---|---|---|---|
| $P = 0.1$ | 0.0500 | 0.0500 | 0.0500 | 0.0500 | 0.0500 | 0.0500 | 0.0500 | 0.0500 | 0.0500 | 0.5500 |
| $P = 0.3$ | 0.0500 | 0.0500 | 0.0500 | 0.0500 | 0.0500 | 0.0500 | 0.0500 | 0.0500 | 0.0500 | 0.5498 |
| $P = 0.5$ | 0.0500 | 0.0500 | 0.0500 | 0.0500 | 0.0500 | 0.0500 | 0.0500 | 0.0500 | 0.0500 | 0.5499 |
| $P = 0.7$ | 0.0500 | 0.0500 | 0.0500 | 0.0501 | 0.0500 | 0.0501 | 0.0501 | 0.0500 | 0.0503 | 0.5494 |
| $P = 0.9$ | 0.0500 | 0.0500 | 0.0514 | 0.0504 | 0.0500 | 0.0539 | 0.0500 | 0.0507 | 0.0506 | 0.5430 |

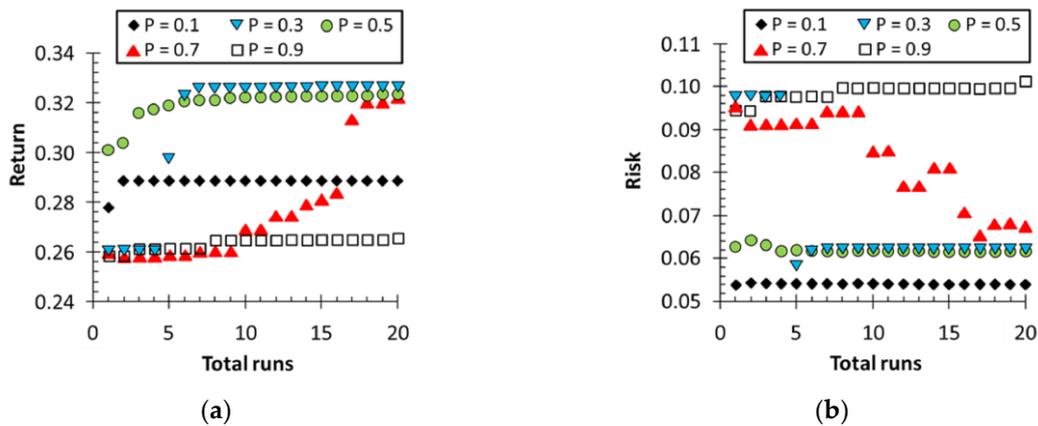

(a)     (b)

**Figure 2.** (a) Return and (b) risk values for each run of the solution algorithm for the decoupled Kelly model at various values of risk parameter $P$.

Table 4. Final portfolios generated for the decoupled Kelly model.

|  | $F_1$ | $F_2$ | $F_3$ | $F_4$ | $F_5$ | $F_6$ | $F_7$ | $F_8$ | $F_9$ | $F_{10}$ |
|---|---|---|---|---|---|---|---|---|---|---|
| $P = 0.1$ | 0.0500 | 0.0500 | 0.0500 | 0.0501 | 0.1438 | 0.0500 | 0.0502 | 0.0532 | 0.0500 | 0.4339 |
| $P = 0.3$ | 0.0500 | 0.0500 | 0.0500 | 0.0500 | 0.0501 | 0.0500 | 0.0500 | 0.0501 | 0.0500 | 0.5498 |
| $P = 0.5$ | 0.0501 | 0.0500 | 0.0503 | 0.0503 | 0.0509 | 0.0500 | 0.0505 | 0.0638 | 0.0500 | 0.5341 |
| $P = 0.7$ | 0.0522 | 0.0517 | 0.0720 | 0.0544 | 0.0519 | 0.0698 | 0.0500 | 0.0608 | 0.0500 | 0.5039 |
| $P = 0.9$ | 0.0500 | 0.0500 | 0.1451 | 0.1045 | 0.0500 | 0.1182 | 0.0939 | 0.1096 | 0.1143 | 0.1840 |

The results in figure 1 show that the risk and return functions in the MV model converge to the same values for all values of risk parameter. A typical run requires ~1000's of iterations and lasts ~100 seconds. The results in table 3 show that the solver generates the same portfolio at all values of risk parameter. The solver generates a result that is biased towards stock 10 due to its large drift-to-volatility and Sharpe ratios. The Kelly model, in comparison, converges to the same portfolio as the MV model for only 3 of the 5 values of risk parameter used in this study. The convergence rate in the decoupled Kelly model for $P = 0.3$ and $P = 0.5$ is similar to the convergence rate in the MV model. For $P = 0.7$, the Kelly model eventually converges to the same portfolio but the convergence rate is significantly slower.

For $P = 0.1$, the decoupled Kelly model converges to a different portfolio than the MV model at a fast rate. The portfolio return and risk are both lower for $P = 0.1$, which is to be expected as risk and return should scale in proportion. The Kelly model returns a more diversified portfolio than the MV model for $P = 0.1$. This would also be expected for low values of risk parameter, which are meant to represent an investor's aversion to risk.

When the risk parameter is set to $P = 0.9$, the decoupled Kelly model returns a counter-intuitive result. Returns are lower and risk is higher. This is indicative of either mis-convergence or simply a very slow convergence rate. In the case of mis-convergence, it is possible that the solution algorithm gets stuck at a local maximum in the decoupled Kelly model for $P = 0.9$.

To validate the accuracy of the objective functions used in each model, the return to risk ratios were calculated for the portfolios using a Monte Carlo simulation. $10^4$ samples were taken for each asset. These results allow for a direct comparison of the ratios predicted by the objective functions in equations (16) and (17). These results are shown in figure 3.

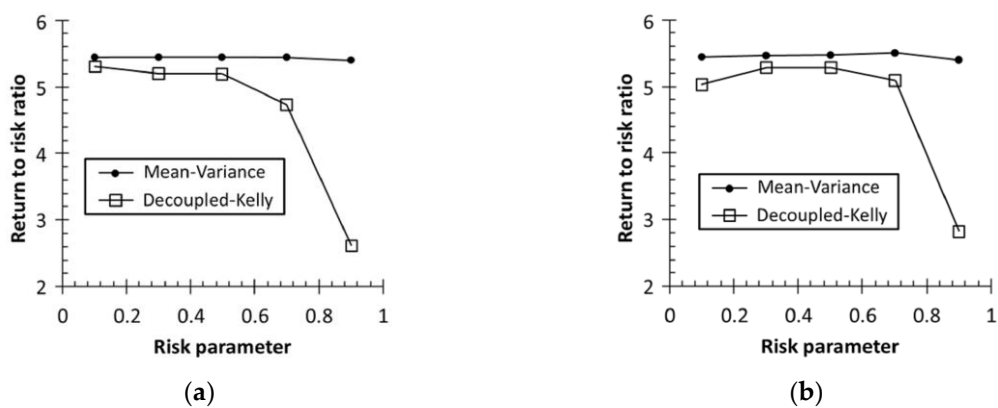

(**a**)  (**b**)

**Figure 3.** (a) Return-to-risk ratios for the portfolios generated from differential evolution for each model. Ratios are shown as a function of risk parameter. (b) Return-to-risk ratios generated using Monte Carlo for the portfolios used in (a).

The pairs of curves shown in figure 3(a) and 3(b) are very similar and the calculations of return and risk from the objective functions are consistent. The curves in each graph follow the same trend. The Monte Carlo model also shows the anomalous very low value for risk-to-return ratio in the decoupled Kelly model for $P = 0.9$. These results confirm the accuracy of the returns and risk predicted under the decoupled Kelly model.

**5. Discussion and conclusions**

Although the algorithm can return what is known to be the optimized solution for both models (depending on the value of risk parameter), there is a clear difference in the convergence rates. To improve the convergence rate in the decoupled Kelly model, the mutation step in the solution algorithm needs to be modified. A number of mutation strategies can be found in (Qin, et. al. 2009). It is well documented in the differential evolution literature that different objective functions have different responsivity to different mutation strategies. A mutation strategy may need to be tailored specifically to the decoupled Kelly problem for different values of risk parameter.

In conclusion, the results in this paper show how Kelly's Criterion can be implemented into a portfolio optimization model that combines risk and return into a single objective function using a risk parameter. The two models tested in this study return the optimized portfolios for moderate values of risk parameter, however, the returns predicted by the two portfolios are slightly different. The solutions to these models were found using a differential evolution algorithm. The rate of convergence of the algorithm was slower for the decoupled Kelly model than for the MV model. It is possible that the decoupled Kelly model mis-converges for one of the values of risk parameter tested in this study. The accuracy of the objective functions in these models was confirmed using a Monte Carlo simulation that calculated return-to-risk ratios using the portfolios generated from the optimization problem. The results are a proof of concept and show that the decoupled Kelly model is a valid portfolio optimization model that can produce an optimized portfolio as well as accurate calculations of return and risk. Further work in this area should focus on improving the solution algorithm to more effectively and quickly reach the solution to the decoupled Kelly problem.

**Acknowledgments:** The author would like to acknowledge many useful conversations with Joe Gittings and for his help with developing the solution algorithm, as well as useful conversations with Ralph Vince and Erik Shamsud-Din.

**Declaration of Interest:** The author reports no conflicts of interest. The author alone is responsible for the content and writing of the paper.